\documentstyle[prl,preprint,eqsecnum,aps,epsf]{revtex}
\draft
\begin{document}
\title
{Instantonic approach to triple well potential}
\author
{Soo-Young Lee$^a$, Jae-Rok Kahng$^a$, Sahng-Kyoon Yoo$^b$, D.K.Park$^c$, 
C.H.Lee$^d$,\\ 
Chang Soo Park$^e$,
Eui-Soon Yim$^f$}
\address
{$^a$ Department of Physics, College of Science, Korea University,
Seoul 136-701, Korea \\
$^b$ Department of Physics, Seonam University, Namwon, Chunbuk 590-170, Korea \\
$^c$ Department of Physics, Kyungnam University, Masan 631-701, Korea \\
$^d$ D\&S Dept.,R\&D Center, Anam Industrial CO., LTD, Seoul 133-120, Korea \\ 
$^e$ Department of Physics, Dankook University, Cheonan  330-180, Korea \\
$^f$ Department of Physics, Semyung University, Chechon 390-230, Korea}
\date{\today}
\maketitle
\begin{abstract}
\indent By using a usual instanton method we obtain the energy splitting due to
quantum tunneling through the triple well barrier. It is shown that
the term related to the midpoint of the energy splitting in propagator 
is quite different from that of double well case, in that 
it is proportional to the algebraic average of the frequencies of
the left and central wells.
\end{abstract}


\section{Introduction}
The Euclidean time formalism of path integral has been applied successfully
to the problems in which quantum tunneling occurs\cite{Cole79,Raja87}. 
In most cases, their results are
consistent with the WKB approximation\cite{Gild77} as well as 
numerical calculations\cite{Bane78}.
Recently much attention has been paid to periodic instanton 
method\cite{Khle91,Lian94}
which can be applied to high-energy quantum tunneling and has a possibility to
explore the baryon- and lepton-number violation in electroweak theory.
Furthermore, there is another approach called valley instanton method that 
allows exploration of the quantum tunneling from false to true 
vacuua\cite{Aoya96,Aoyahep}.
\\
\indent
In general  the Euclidean time formalism gives the information about the ground
state, i.e., the ground state wavefunction and the ground state energy.
In the case of double well, it allows applications of path integral by
opening a possiblity of finding a classical path in Euclidean space,
and explains quantum tunneling
through the barrier between wells resulting in an energy splitting of 
the degenerate ground states.
This energy splitting was also investigated through the WKB approximation and
numerical calculation as well as through the path integral with 
Euclidean time(instanton
method).
These results are consistent with each other and it was conjectured 
that the instanton method
may give more exact result than the WKB approximation[3].
\\
\indent
We here consider a triple well potential on the ground of the instanton method.
For simplicity, we take the triple well potential in the following form;
\begin{equation}
V(\phi)=\alpha \phi^2 ( \phi - \beta )^2 (\phi + \beta )^2.
\label{potential}
\end{equation}
This potential has three wells among which two side wells have 
topologically identical vacuua,
and central well is rather broader as shown in Fig. 1. 
This potential form may be
realized by designing supperlattices with different compound compositions
whose applications might  come true in electronic devices such as 
laser without inversion\cite{Harr89}.
Therefore, understanding the structure of the energy level, which influences
the dynamics of electrons,  is essential in constructing electronic devices.
\\
\indent
Compared with the double well case, there are two different aspects.
Firstly, the energy spectrum forms discrete blocks, each of which is composed
of three levels. The level located in the middle of each block 
has different symmetry; for example,
energy levels in the lowest block is arranged in the order of 
even, odd and even states in energy.
Secondly, the central vacuum is not identical with the others locating at 
the sides.
This indicates that the structure of the block can be adjusted by changing the 
curvature of the central
vacuum. This aspect might be important in practical electronic engineering.
\\
\indent
In this paper we obtain the central position and width of 
the lowest block, i.e., energy splitting, and
partial information about the wavefunctions of the lowest block 
using the instanton method.
The term related to the midpoint of energy splitting turns out 
to be proportional to algebraic
average of the frequencies of the left and central wells.
In Sec.II, we find out what physical quantities are related to 
the propagator with large
Euclidean time. One instanton contribution to the propagator 
is calculated in Sec.III,
and in order to obtain a desired form of the propagator we use the 
dilute gas approximation
for the configurations composed of multi-instantons and anti-instantons 
in Sec.IV.
In Sec.V a brief conclusion is given.
Throughout this paper, we take $\hbar =1$ and the mass $m=1$ for simplicity.
\section{Propagator at large Euclidean time}
We consider a triple well problem in which barriers are so high that at least
one discrete eigenvalue can exist below the barrier height. 
Generally in the case of
triple well potential, an energy level is split into three levels with slightly
different energy values due to the quantum tunneling through barriers.\\
\indent
Let $|L_0>$, $|C_0>$, and $|R_0>$ denote the normalized lowest
energy eigenstates of the isolated left,
central, and right well, repectively. When barriers are appropriately 
high and wide, the three lowest eigenstates
in triple well can be obtained as following through a  variational
method\cite{Ribo},
\begin{eqnarray}
|E'_0> &\simeq& 
\frac{1}{\sqrt{2+a_+^2}} ( |L_0>+ a_+ |C_0> + |R_0> )\label{state1} \\
|E'_1> &\simeq& \frac{1}{\sqrt{2}} ( |L_0> - |R_0> )  \label{stae2}\\
|E'_2> &\simeq& \frac{1}{\sqrt{2+a_-^2}} ( |L_0>- a_- |C_0> + |R_0> )
\label{state3}
\end{eqnarray}
where
\begin{equation}
a_\pm \equiv \frac{1}{2} \left [ \pm \frac{H_{LL}- H_{CC}}{|H_{LC}|} +
\sqrt{(\frac{H_{LL}- H_{CC}}{|H_{LC}|})^2 + 8} \right ].
\end{equation}
Here,
\begin{equation}
H_{LL}\equiv <L_0|H|L_0>,
 ~~~H_{CC}\equiv <C_0|H|C_0>, ~~~H_{LC}\equiv <L_0|H|C_0>,
\end{equation}
where $H$ is the total Hamiltonian.
In getting these states, we use the fact that the ground state 
wavefunctions of each well are taken to be almost separated due to 
large potential barriers and
there is a two-fold symmetry.
The lowest three eigenstates in Eqs.(\ref{state1}-\ref{state3}) are 
orthogonal to each other and
 have following eigenvalues:
\begin{eqnarray}
<E'_0| H |E'_0>\equiv E'_0 &\simeq& 
\frac{2H_{LL}+a_+^2 H_{CC}- 4a_+ |H_{LC}|}{2+a_+^2} \\
<E'_1| H |E'_1>\equiv E'_1 &\simeq& H_{LL} \\
<E'_2| H |E'_2>\equiv E'_2 &\simeq& 
\frac{2H_{LL}+a_-^2 H_{CC}+ 4a_-|H_{LC}|}{2+a_-^2}.
\end{eqnarray}
If the curvatures of the potential with three vacuua are equivalent, then
this problem is reduced to three-fold degenerate one. 
In this case, $H_{LL}=H_{CC}$
and $a_\pm = \sqrt{2}$. Thus the lowest energy eigenstate is split
into three eigenstates having equal spacing
 $\Delta E = \sqrt{2} |H_{LC}|$ between the eigenvalues. \\
\indent
It is well known that for a large Euclidean time interval the propagator gives
the information of the ground state energy. In the triple well case, 
the propagator from left($\phi_i$) to central vacuum($\phi_f$) is approximately
described as follows:
\begin{equation}
<\phi_f | e^{-2HT} | \phi_i> 
\simeq \sum_{j=0}^{2}<\phi_f | E'_j><E'_j |\phi_i > e^{-2E'_j T}.
\label{van}
\end{equation}
 Here, since the  wave function of the state $|E'_1>$ is odd , $<\phi_f | E'_1>$
vanishes, and the coefficients of the terms corresponding to 
$j = 0$ and $j = 2$ in the Eq.(\ref{van}) are equal,
that is,
\begin{equation}
\frac{2 a_+}{2+a_+^2} = \frac{2 a_-}{2+a_-^2}= 
\frac{2}{\sqrt{(\frac{H_{LL}-H_{CC}}{\mid H_{LC}\mid } )^2 + 8}}.
\end{equation}
Then, the propagator becomes 
\begin{equation}
<\phi_f | e^{-2HT} | \phi_i> 
\simeq \frac{2}{\sqrt{(\frac{H_{LL}-H_{CC}}{\mid H_{LC}\mid } )^2 + 8}} 
<\phi_f | C_0><L_0 |\phi_i > e^{-2E T} \sinh 2 \Delta E T,
\label{kernel}
\end{equation}
where
\begin{equation}
E \equiv \frac{E'_0+E'_2}{2}, ~~~~ \Delta E \equiv \frac{E'_2 - E'_0}{2}.
\end{equation}
This result shows that, in the case of triple well, the propagator 
for large Euclidean
time $T$ tells us about the width and the central position of 
the lowest energy block.
\\
\indent
In the following sections we will calculate the propagator by 
using the instanton method,
and show that the propagator with the dilute instanton gas approximation
takes the same form as Eq.(\ref{kernel}).
\section{One-instanton contribution}
In the present section we will calculate the one-instanton contribution to
the propagator.  \\
The Euclidean propagator from ($\phi_i,-T$) to ($\phi_f,T$) is written as
\begin{equation}
<\phi_f | e^{-2HT} | \phi_i> = 
\int_{\phi_i,-T}^{\phi_f,T} D \phi e^{-S_E [\phi ]},
\label{propa}
\end{equation}
where the Euclidean action is
\begin{equation}
S_E [\phi ] = 
\int_{-T}^{T} d\tau \left[ \frac{1}{2} \dot{\phi}^2 + V(\phi) \right].
\label{action}
\end{equation}
The path giving major contribution is determined by the classical 
equation of motion, i.e.,
$\delta S_E /\delta \phi = 0 $.
For the triple well potential in Eq.(1.1) whose central well is wider than
the side wells, the equation of motion becomes
\begin{equation}
\ddot{\phi} = \frac{\partial V(\phi )}{\partial \phi }
 = 2 \alpha \phi ( 3 \phi^4 - 4 \beta^2 \phi^2 + \beta^4 ).
\end{equation}
The classical solution of this equation of motion, so-called instanton, is
\begin{equation}
\phi_{cl}(\tau) = 
\frac{-\beta }{\sqrt{1 + e^{2\beta^2 \sqrt{2\alpha } (\tau - \tau_0)}}} .
\end{equation}
We here take a boundary condition as $\phi_i=-\beta$ 
and $\phi_f=0$ (from left to central vacuum).
\indent
Substituting the solution into Eq.(\ref{action}),  and taking 
a $T \rightarrow \infty$ limit,
 we obtain
\begin{equation}
\lim_{T \rightarrow \infty} S_E [\phi_{cl}] = \frac{\sqrt{2\alpha} \beta^4}{4}.
\end{equation}
\indent In order to evaluate the contribution of paths neighboring 
on the instanton solution to 
the propagator, we define
\begin{equation}
\phi (\tau ) = \phi_{cl}(\tau ) + \eta (\tau ),
\end{equation}
where $\eta(\tau )$ is an arbitrary function satisfying a boundary condition:
\begin{equation}
\eta (-T) =  \eta (T) = 0.
\end{equation}
Now the action can be divided into the two parts: one from instanton
and the other from the 
paths neighboring  on the instanton. Therefore, the total action becomes
\begin{equation}
S_E = S_{E}[\phi_{cl}] + \int_{-T}^{T} d\tau \left[ \frac{1}{2} \dot{\eta}^2 +
                    \alpha Y(\tau ) \eta ^2 \right],
\label{eta}
\end{equation}
where
\begin{equation}
Y(\tau ) \equiv 15 \phi_{cl}^4(\tau ) - 12 \beta^2 \phi_{cl}^2(\tau ) + \beta^4.
\end{equation}
In Eq.(\ref{eta}), we keep up to the second order terms of $\eta $.
Then the propagator can be written as
\begin{equation}
<0 | e^{-2HT} | -\beta> = e^{-S_{E}[\phi_{cl}]} I(T),
\label{pro}
\end{equation}
where
\begin{equation}
I(T)= \int_{(-T,0)}^{(T,0)} D \eta e^{-\int d\tau \eta \hat{M} \eta },
\end{equation}
and
\begin{equation}
\hat{M} = -\frac{1}{2} \frac{d^2 }{d \tau^2 } + \alpha Y(\tau ).
\end{equation}

Our next task is to calculate the functional integral $I(T)$ which 
has approximated
to a Gaussian form. This is, however, not simple because it contains 
a zero mode, which
is easily seen from the fact,
\begin{equation}
\hat{M} \dot{\phi }_{cl}(\tau ) = 0.
\end{equation}
It indicates that $\dot{\phi}_{cl}(\tau )$ is the eigenvector 
corresponding to zero
eigenvalue of $\hat{M}$. Therefore, the zero mode should be treated 
carefully in calculating $I(T)$.
Here we follow the method used by Gildener and Patrascioiu\cite{Gild77} 
in performing
the functional integral. 
\\
\indent
Let the eigenfunctions and corresponding eigenvalues of $\hat{M}$ 
be $\psi_m$ and $\varepsilon_m$, respectively,
\begin{equation}
\hat{M} \psi_m = \varepsilon_m \psi_m.
\end{equation}
Then the small fluctuating path  $\eta(\tau )$ can be expanded as a linear
combination
of the eigenfunctions:
\begin{equation}
\eta (\tau ) = \sum_m c_m \psi_m (\tau ).
\end{equation}
In the Gaussian functional integral $I(\tau )$, 
the integral variable $\eta$ can now
be changed to the coefficients $c_m$, which gives a formal solution including
the Jacobian in the form
\begin{eqnarray}
I(T)&=& \mid \frac{\partial \eta}{\partial c_n} \mid 
\int \prod_m d c_m e^{-\sum_m \varepsilon_m c_m^2 }  \nonumber \\
&=& \mid \frac{\partial \eta}{\partial c_n} \mid 
\prod_{m=0}^{\infty} \sqrt{\frac{\pi}{\varepsilon_m}}.
\label{formal}
\end{eqnarray}
To surmount the difficulty of the zero mode, one can use the 
collective coordinate method,
 which gives
\begin{equation}
I(T)=2T\sqrt{S_E [\phi_{cl} ]} I_0,
\label{IT}
\end{equation}
where
\begin{equation}
I_0 \equiv \mid \frac{\partial \eta}{\partial c_n} \mid 
\prod_{m=1}^{\infty} \sqrt{\frac{\pi}{\varepsilon_m}} .
\label{IT0}
\end{equation}
\indent
In order to find an explicit expression for $I_0$ we use the method
of changing variables\cite{Dash74}.
Then the functional integral $I(T)$ can be represented as
\begin{equation}
I(T)=
\left( 2\pi N(T) N(-T) \int_{-T}^{T} \frac{d\tau}{N^2(\tau )} \right)^{-1/2},
\end{equation}
where
\begin{equation}
N(\tau )=\dot{\phi}_{cl} =  
\frac{\beta^3 \sqrt{2\alpha}  
e^{2\beta^2 \sqrt{2\alpha} \tau}}{(1+e^{2\beta^2\sqrt{2\alpha} \tau })^{3/2}}.
\end{equation}
For large $T$, this becomes
\begin{equation}
I (T) = \beta \left( \frac{2}{\pi} \sqrt{2\alpha} \right)^{1/2}
e^{-\frac{1}{2} \beta^2 \sqrt{2\alpha} T}.
\label{changing}
\end{equation}
Comparing Eq.(\ref{changing}) with Eq.(\ref{formal}) and Eq.(\ref{IT0}), 
we obtain
\begin{equation}
I_0 = \frac{\beta}{\pi} (2 \beta^2 \sqrt{2\alpha} \varepsilon_0)^{1/2}
e^{-\frac{1}{2} \beta^2 \sqrt{2\alpha} T}.
\label{I0}
\end{equation}
Imposing the boundary condition, $\psi_n(-T)=\psi_n(T)=0$, yields 
the lowest energy
$\varepsilon_0$ in WKB approximation at large $T$[3] as
\begin{equation}
\varepsilon_0 = 8 \alpha \beta^4 e^{-2\beta^2 \sqrt{2\alpha} T}.
\label{epsilon0}
\end{equation}
Inserting this into Eq.(\ref{I0}), we get
\begin{equation}
I_0 = \frac{4\beta^3}{\pi}(\alpha \sqrt{2\alpha} )^{1/2}
e^{-\frac{3}{2} \beta^2 \sqrt{2\alpha} T}.
\end{equation}
From Eq.(\ref{IT}), therefore, the final form of the functional integral $I(T)$
is expressed in the form 
\begin{equation}
I(T)=\frac{8}{\pi}\beta^3 T\sqrt{S_E[\phi_{cl}]} (\alpha \sqrt{2\alpha})^{1/2}
e^{-\frac{3}{2} \beta^2 \sqrt{2\alpha} T}.
\end{equation}
In the case of double well, it is well known that the exponent 
in the exponential function
is closely related to the curvature of vacuum, that is, if the 
second derivative of the
double well potential at vacuum positions is $\omega^2$, then the 
exponential function
is given by $e^{-\omega T}$. However, in our triple well potential 
the curvature of
the side vacuua is different from that of the central vacuum (Fig.1). Note that
in this case the exponent is also related to the potential curvatures 
at the vacuum positions in a way of
algebraic averaging, that is,
\begin{equation}
\omega \equiv \frac{\omega_1 + \omega_2 }{2} = 
\frac{3}{2} \beta^2 \sqrt{2\alpha}
\end{equation}
where
\begin{equation}
\omega_1^2 \equiv V''(-\beta ) 
= 8 \alpha \beta^4,~~~~  \omega_2^2 \equiv V''(0)
= 2\alpha \beta^4.
\end{equation}
As will be seen shortly, $\omega /2$ is the central position of the energy 
block composed
of three almost degenerate lowest levels.
\\
\indent
From Eq.(\ref{pro}), the contribution of one instanton to 
the propagator becomes
\begin{equation}
<0 | e^{-2HT} | -\beta> 
=2\kappa T \sqrt{S_E[\phi_{cl}]}\sqrt{\frac{\omega}{\pi}} e^{-S_{E}[\phi_{cl}]}
 e^{-\omega T},
\end{equation}
where
\begin{equation}
\kappa \equiv 4 \beta^4 \sqrt{\frac{2 \alpha }{3\pi }}.
\end{equation}
In the next section the contribution of multi-instanton to the 
propagator will be explored.
\section{Multi-instanton contribution}
In the previous section we consider the contribution of the 
stationary solution(instanton) and
its neighboring paths. However, since the path integral 
in Eq.(\ref{propa}) includes all possible
paths starting from ($-\beta, -T$) and ending at ($0,T$), 
let us consider the paths consisting of $n_1$ instantons and
$n_2$ anti-instantons which are located at large distance with each 
other in Euclidean time so that the interaction
between them can be ignored.
In order to satisfy the boundary condition, the number of instanton 
must be one larger
than that of anti-instanton. Furthermore, since the total number of 
possible configurations
is $2^{n_2}$, the multi-instanton contribution to the propagator is
\begin{eqnarray}
& & <0 | e^{-2HT} | -\beta > \nonumber \\
&&=\sum_{n_1=1}^{\infty} \sum_{n_2=0}^{\infty} \delta_{n_1 -n_2, 1}
\frac{2^{n_2}}{(n_1 + n_2)!} e^{-(n_1 + n_2) S_{E}[\phi_{cl}]}
( 2\kappa T \sqrt{S_E[\phi_{cl}]} )^{n_1 + n_2}  
\sqrt{\frac{\omega}{\pi}}e^{-\omega T} \nonumber \\
&&= 2 \kappa T \sqrt{S_E[\phi_{cl}]}
\sqrt{\frac{\omega}{\pi}} e^{-\omega T} 
e^{-S_{E}[\phi_{cl}]}
\sum_{n=0}^{\infty} 
\frac{[2 e^{-2 S_{E}[\phi_{cl}]} (2 \kappa T \sqrt{S_E[\phi_{cl}]} )^2]^{n}}
{(2n+1)!}.
\label{dil}
\end{eqnarray}
Using the identity,
\begin{equation}
\sum_{n=0}^{\infty} \frac{x^n}{(2n+1)!} = \frac{1}{\sqrt{x}}\sinh \sqrt{x},
\end{equation}
the infinite sum in Eq.(\ref{dil}) is represented by a hyperbolic sine function.
Therefore, the final result is of the following form,
\begin{equation}
<0 | e^{-2HT} |-\beta >
= \sqrt{\frac{\omega}{2\pi}}e^{-\omega T} 
\sinh (2\sqrt{2} \kappa T \sqrt{S_E[\phi_{cl}]}e^{-S_{E}[\phi_{cl}]}
).
\end{equation}
Comparing this result with Eq.(\ref{kernel}), we get the central 
position $E$ and the
width $2\Delta E$ of the lowest energy block in terms of potential 
parameters, $\alpha$ and $\beta$,
which are
\begin{equation}
E = \frac{3}{4} \beta^2 \sqrt{2\alpha}
\end{equation}
and
\begin{equation}
\Delta E = \sqrt{\frac{8}{3\pi}} 
(2\alpha )^{3/4} \beta^4 e^{-\frac{1}{4}\beta^4 \sqrt{2\alpha}}.
\end{equation}
For sufficiently large $\beta$, in other words, both large well 
separation and high
potential barrier, the energy splitting decreases exponentially as expected.
Moreover, the product of the amplitudes of the ground state wavefunction 
at vacuum positions
is expressed as
\begin{equation}
<0 | E'_0><E'_0 |-\beta>
=\frac{\beta}{4}\left( \frac{3}{\pi}\sqrt{2\alpha} \right)^{1/2}.
\end{equation}
\section{Conclusion}
The quantum tunneling for the triple well potential case is 
investigated by means of
the instanton method. Unlike the case of double well potential, the 
curvature of the
central well is topologically different from those of the others, which results
in the fact that the midpoint of energy splitting is proportional to the 
algebraic
average of the frequencies of the left and central wells.
We also obtained the energy splitting $\Delta E$ of the lowest
block. Although this kind of approach
can be extended to the $n$-tuple($n >3$) well, much complication in 
counting the
possible configuration gets involved. In the case of quadruple 
well, for example, two
different types of instanton and anti-instanton solutions should 
be considered carefully.
The study about this subject is in progress.
\begin{center}
 ACKNOWLEDGEMENT 

This work was supported in part by Nondirected Research Fund, Korea Research 
Foundation, 1995 and by Korea Science and Engineering Foundation
(961-0201-005-1).
\end{center}

\begin{figure}
\caption{Triple well potential as a function of position.}
\end{figure}

\newpage
\epsfysize=20cm \epsfbox{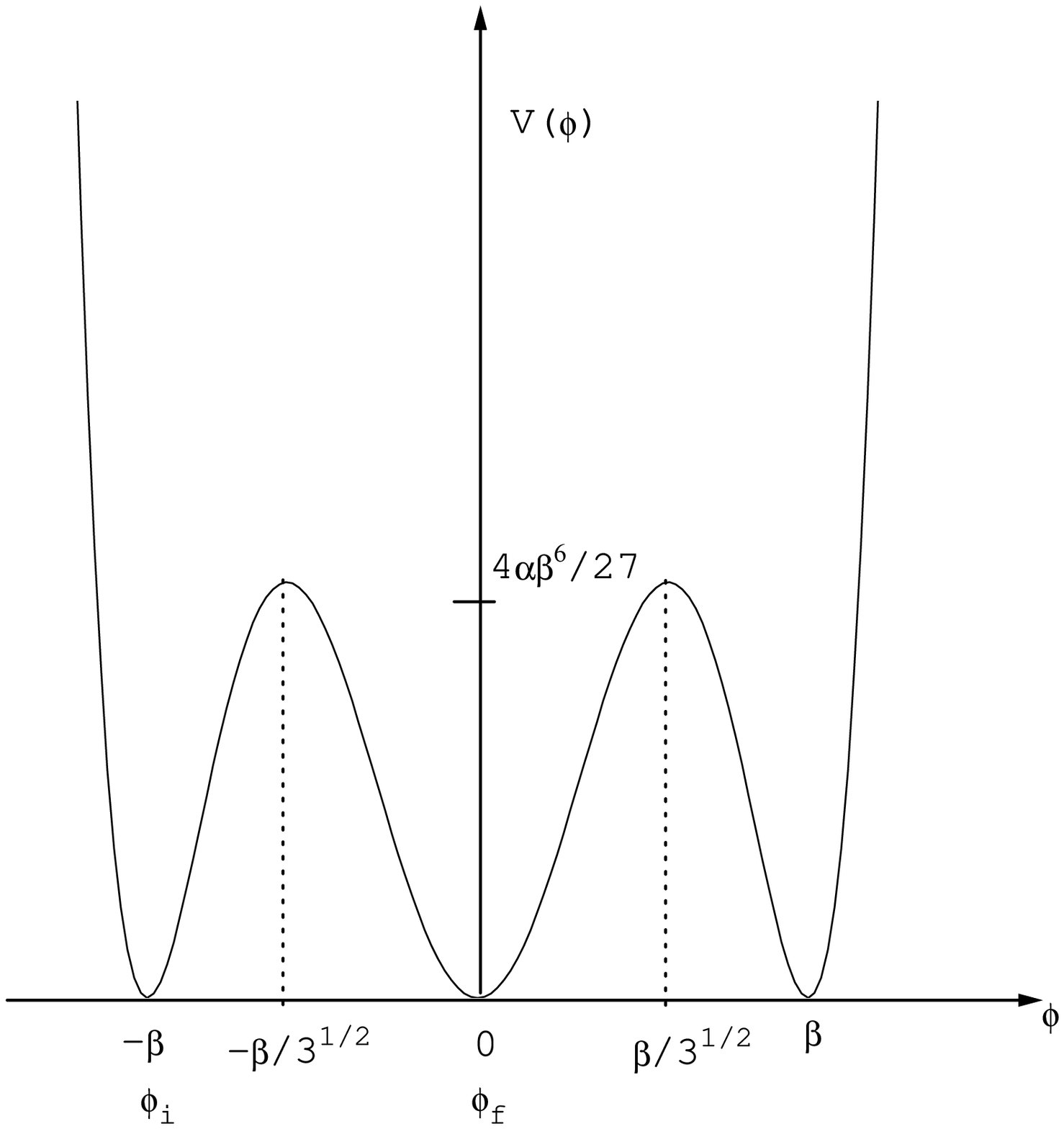}


\normalsize
\begin{thebibliography}{99}
\bibitem{Cole79} S.Coleman, "The uses of Instantons",
in {\it The why's of subnuclear physics}, 
ed. A.Zichichi (Plenum, New York, 1979).
\bibitem{Raja87} R.Rajaraman, {\it Solitons and Instantons}, (North-Holland, the Netherlands, 1987).
\bibitem{Gild77} E.Gildener and A.Patrascioiu, Phys. Rev. D{\bf 16}, 423 (1977).
\bibitem{Bane78} K.Banerjee and S.P.Bhatnagar, Phys. Rev. D{\bf 18}, 4767 (1978).
\bibitem{Khle91} S.Yu.Khlebnikov, V.A.Rubakov, and P.G.Tinyakov, Nucl. Phys. B{\bf 367}, 334 (1991).
\bibitem{Lian94} J.Q.Liang and H.J.W.M\"{u}ller-Kirsten, Phys. Rev. D{\bf 50}, 6519 (1994);
{\bf 51}, 718 (1995).
\bibitem{Aoya96} H.Aoyama, T.Harano, M.Sato, and S.Wada, Nucl. Phys. B{\bf 466}, 127 (1996).
\bibitem{Aoyahep} H.Aoyama, T.Harano, H.Kikuchi, M.Sato, and S.Wada, hep-th/9606159.
\bibitem{Harr89} S.E.Harris, Phys. Rev. Lett. {\bf 62}, 1033 (1989).
\bibitem{Ribo} R.L.Liboff, 
{\it Introductory Quantum Mechanics}, (Addison-Wesley Publishing Company, Inc. 1992).
\bibitem{Dash74} R.F.Dashen, B.Hasslacher, and A.Neveu, Phys. Rev. D{\bf 10}, 4114 (1974).
\end{thebibliography}
\end{document}